\documentclass[pdflatex,sn-mathphys-num]{sn-jnl}

\usepackage{graphicx}%
\usepackage{multirow}%
\usepackage{amsmath,amssymb,amsfonts}%
\usepackage{amsthm}%
\usepackage{mathrsfs}%
\usepackage[title]{appendix}%
\usepackage{xcolor}%
\usepackage{textcomp}%
\usepackage{manyfoot}%
\usepackage{booktabs}%
\usepackage{algorithm}%
\usepackage{algorithmicx}%
\usepackage{algpseudocode}%
\usepackage{listings}%
\theoremstyle{thmstyleone}%
\theoremstyle{thmstyletwo}%

\theoremstyle{thmstylethree}%

\begin{document}

\title[Article Title]{AI models of unstable flow exhibit hallucination}


\author{\fnm{Ramdhan} \sur{Wibawa}}

\author*{\fnm{Birendra} \sur{Jha}}\email{bjha@usc.edu}

\affil{\orgdiv{Department of Chemical Engineering and Materials Science}, \orgname{University of Southern California}, \orgaddress{\street{925 Bloom Walk}, \city{Los Angeles}, \postcode{90089}, \state{CA}, \country{USA}}}


\abstract{
We report the first systematic evidence of hallucination in AI models of fluid dynamics, demonstrated in the canonical problem of hydrodynamically unstable transport known as viscous fingering. AI-based modeling of flow with instabilities remains challenging because rapidly evolving, multiscale fingering patterns are difficult to resolve accurately. We identify solutions that appear visually realistic yet are physically implausible, analogous to hallucinations in large language models. These hallucinations manifest as spurious fluid interfaces and reverse diffusion that violate conservation laws. We show that their origin lies in the spectral bias of AI models, which becomes dominant at high flow rates and viscosity contrasts. Guided by this insight, we introduce DeepFingers, a new framework for AI-driven fluid dynamics that enforces balanced learning across the full spectrum of spatial modes by combining the Fourier Neural Operator with a Deep Operator Network to predict the spatiotemporal evolution of viscous fingers. By conditioning on both time and viscosity contrast, DeepFingers learns mappings between successive concentration fields across regimes. The framework accurately captures tip splitting, finger merging, and channel formation while preserving global metrics of mixing. The results open a new research direction to investigate fundamental limitations in AI models of physical systems.
}


\maketitle

\section*{Introduction}\label{sec1}

Hydrodynamic instabilities such as viscous fingering influence many natural and engineered systems, including chemical~\cite{losey01}, pharmaceutical~\cite{dunn00,baldyga10}, food processing~\cite{cullen09}, mantle convection~\citep{olson90}, and groundwater systems~\cite{dentz2011JCH}.
Viscous fingering  arises at the interface between two fluids of differing viscosities when a less viscous fluid displaces a more viscous fluid within a porous domain~\cite{homsy1987,araktingi1993viscous,jha2011fluid,zheng2015controlling,pinilla2021experimental}. This classical fluid mechanics phenomenon, primarily driven by the Saffman–Taylor instability, has been extensively studied in both miscible and immiscible fluid systems~\cite{yazdi2018numerical,chui2015interface,moyles2015fingering}. In miscible flow displacements, such as water displacing glycerin in Hele-Shaw cells~\cite{bacri92,petitjeans1996miscible} or CO$_2$ displacing crude oil or water in petroleum or geothermal reservoirs~\cite{ren2025_CO2water_NMR,park1984two,zheng2015controlling}, the advancing interface evolves into intricate finger-like patterns that grow increasingly complex as the instability develops throughout the domain.
Accurate prediction of such instabilities is crucial for numerous applications, including enhanced oil recovery, CO$_2$ sequestration, groundwater remediation, hydraulic fracturing, and the design of microfluidic systems in biomedical engineering~\cite{zheng2015controlling,gao2019active,escala2021bottom,tranjha2020,tu2022facile_VF}. Uncontrolled finger propagation can significantly degrade operational performance by prolonging contaminant removal, reducing reservoir sweep efficiency~\cite{nicolaides2015impact}, and increasing the amount of CO$_2$ required for effective sequestration~\cite{wang2013experimental,gooya2019unstable,berg2012stability}. On the other hand, enhancement of fingering has been proposed as a mechanism for faster mixing under laminar flow conditions~\cite{jha2011fluid,jha2013PRL2013}. Recently, the complexity and randomness of the fingers have been utilized to propose an anti-counterfeiting solution~\cite{zhao2025_anticounterfeit}. 

Predicting viscous fingering is challenging because small perturbations can rapidly evolve into complex flow patterns across multiple scales. Numerical simulation of viscous fingering 
presents notable challenges due to the nonlinear and multiscale nature of the phenomenon~\cite{tan1988simulation,jha2013PRL2013}. The system is governed by coupled partial differential equations (PDEs) derived from the conservation of mass and momentum. These equations link physical parameters such as fluid viscosity and density to the evolving state variables like the fluid concentration field. Through non-dimensionalization, key dimensionless groups such as the viscosity ratio $M$ and the Peclet number Pe emerge to characterize the dynamics of the instability.
Solving these PDEs over large domains (equivalently, large Pe) 
and large viscosity ratios often relies on direct numerical simulation (DNS) techniques, commonly employing finite volume or finite element methods for spatial discretization~\cite{casademunt2004viscous}, along with time-stepping schemes such as Euler integration. However, due to the sensitivity of the flow to perturbations in medium properties and initial conditions and the complex feedback between advection and diffusion, many simulations diverge from experimental or field-scale observations, particularly as fingering patterns evolve into highly intricate structures. Recent progress in AI modeling of fluid flow, especially deep learning models, offers  promise in addressing these challenges. 
This study shows that deep learning models may produce visually convincing predictions that nonetheless violate fundamental physical principles, a failure mode analogous to hallucinations in language models. 

Hallucination of AI models has been extensively documented and studied in large language models—prompting major efforts to understand its societal impact and develop mitigation strategies~\cite{lee2018hallucinations}. However, it has been largely assumed that physics-informed or data-driven AI models operating in scientific domains are immune to such failures. We  report the first systematic evidence of hallucination in AI models of fluid dynamics. Using modern architectures including Vision Transformers, we demonstrate that AI models trained to predict viscous fingering dynamics can exhibit physically inconsistent behavior despite maintaining visual coherence. We further identify the origin of hallucination in fluid-dynamics AI models as spectral bias, wherein learning architectures disproportionately favor certain length scales at the expense of others.

Guided by this insight, we introduce a new deep learning framework, combining DeepONet~\cite{lu2019deeponet} and Fourier Neural Operator (FNO)~\cite{li2020Fourier} frameworks, that enforces balanced learning across the full spectrum of spatial modes. We demonstrate how such a design is necessary to model unstable flows at the full-field scale or, equivalently, at large Peclet numbers representative of real-world systems.
This raises awareness of the need for physical consistency when designing and evaluating deep learning models for scientific applications.



\section*{Results}\label{sec2}

Flows with hydrodynamic instabilities, and viscous fingering in particular, present a multi-fidelity modeling challenge. The bulk fluid region and the advancing fingers of the less viscous fluid evolve dynamically over time, dictated by partial differential equations (PDEs) of mass and momentum balance, while intricate mechanisms emerge at the finger tips and along the interface between the two fluids. These regions exhibit sharp or smooth gradient transitions, depending on the interplay of nonlinear advection and diffusion. Overall spreading and mixing behavior is governed by globally defined flow metrics. Details of the governing PDEs, flow metrics, and their implications for instability are provided in the Supporting Information.
\begin{figure}
    \centering
    \includegraphics[width=\linewidth]{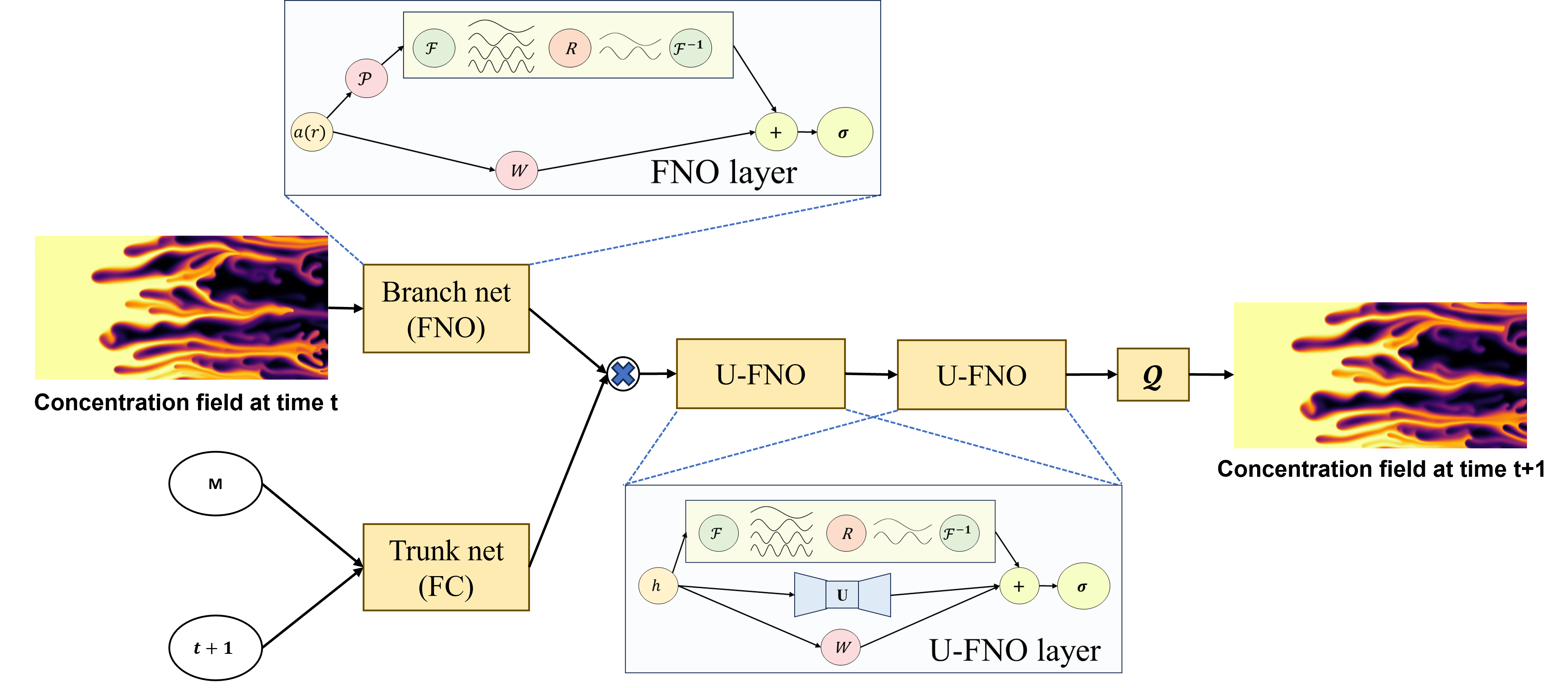}
    \caption{The DeepFingers architecture for modeling viscous fingering of the less viscous fluid (yellow) through a more viscous fluid (black). It applies FNO at the branch network and a fully connected (FC) layer at the trunk network. The outputs are merged before processing in a series of two U-FNO layers and a projection layer ($\mathcal{Q}$) to generate the concentration map at the next time step.}
    \label{fig:FNO-deepOnet}
\end{figure}

On the AI modeling side, it is known that Fourier Neural Operator (FNO) layers, which implement integral kernel operators formulated from Green's function,  provide an  expressive and efficient representation for solving PDEs whose solutions contain multiple length scales. 
The effectiveness of FNO has been demonstrated across a range of PDE benchmarks, where it outperforms other deep learning architectures in both accuracy and computational efficiency~\cite{li2020Fourier,li2023fourier,raonic2023convolutional,liu2025enhancing}. In the broader context of operator learning, DeepONet has also shown strong performance across diverse physical systems~\cite{de2023coupled,michalowska2023don,haghighat2024deeponet,moya2022fed,kontolati2024learning,garg2022variational}, further highlighting the advantages of learning solution operators directly from data. 

Inspired by these developments, we propose a new architecture for modeling flows with fingering instabilities:  DeepFingers. 
The DeepFingers architecture (Fig.~\ref{fig:FNO-deepOnet}) extends the DeepONet framework~\cite{lu2019deeponet} by incorporating a FNO~\cite{li2020Fourier} in the branch network, a fully connected structure in the trunk network, and a series of U-FNO layers~\cite{wen2022u, Zhu_2023} that embed the well-known U-Net~\cite{ronneberger2015u} structure to enhance feature extraction and multi-scale representation within the FNO architecture. The U-FNO layers enhance representational capacity by capturing high-frequency components that are not adequately resolved by the standard Fourier basis~\cite{wen2022u}. 

The framework consists of two separate components: a branch network and a trunk network. The branch network processes the input function at fixed sensors locations, while the trunk network receives the input parameters (often referred to as coordinates or locations in the original formulation). 
This design is well-suited to our objective of investigating the fingering dynamics, where the branch network represents the grid of concentration field and the trunk network process time $t$ and viscosity ratio $M$ as input parameters. 

The assessment of DeepFingers covers four aspects: computational efficiency, hallucination detection, comparison with baseline models using physics-based metrics and spectral analysis, and performance under uncertainty. Predictions are initialized from prescribed initial conditions and advanced auto-regressively in time. Because fingering becomes increasingly unstable as the $M$ value increases, a consequence of the exponential viscosity function (see equations in the Supplementary Information), physically grounded diagnostics and spectral analysis are chosen over pixel-wise errors. 

To evaluate its effectiveness, we compare DeepFingers against two state-of-the-art AI modeling frameworks that employ a two-stage modeling approach, previously developed for cylindrical flow problems~\cite{Geneva2022TransformersFM, beta-vae} and for Darcy flows with small Peclet numbers. 
In addition, we benchmark it against a Vision Transformer (ViT)-based model~\cite{dosovitskiy2020image}, a widely recognized architecture for image-related tasks, used here as a transformer-based baseline. We also note recent progress in applying transformer architectures to physics-based flow modeling, such as the CViT framework~\cite{wang2024cvit}, which further illustrates the growing role of attention mechanisms in scientific machine learning.
DeepFingers is benchmarked against DAE-LSTM and ViT across a range of $M$ values and time horizons to provide a comprehensive measure of model robustness.

Reconstruction of the concentration field from DNS and various deep learning models is presented in Fig.~\ref{fig:concentration}. The comparison includes DeepFingers, ViT, and DAE-LSTM across a range of $M$ values and time. DNS provides a physical benchmark that begins with the emergence of narrow fingers, followed by repeated tip-splitting, merging, and shielding interactions that drive the instability. Accurately reproducing these highly nonlinear and spatio-temporal dynamics directly from data, without solving the governing equations, is an exceptionally difficult task~\cite{pinilla2021experimental}. The proposed DeepFingers demonstrates a strong ability to recover these essential physical characteristics. At low $M$, it correctly predicts short, slowly advancing fingers. As $M$ increases, it captures the accelerated finger growth, more frequent tip-splitting events, and the formation of intricate patterns. At the highest viscosity ratio ($M=55$), the model even reproduces the onset of channeling~\cite{alasker2025scalar}, where high-mobility pathways dominate and significantly alter the fluid displacement process.

\subsection*{Hallucinations in AI models of flows with fingers}
\begin{figure*}
    \centering
    \includegraphics[width=1.0\linewidth]{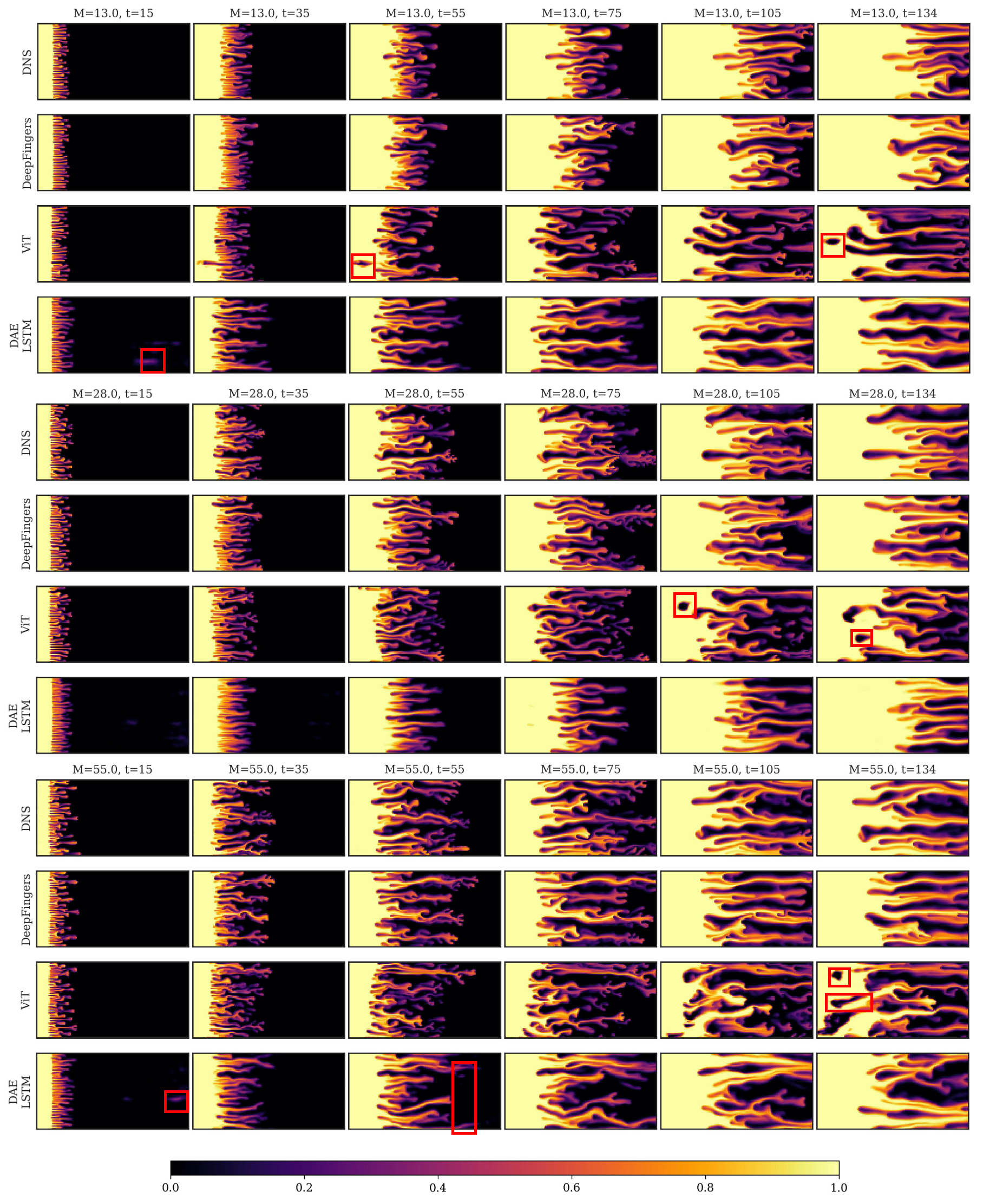}
    \caption{Hallucination detection in AI models of fingering. 
    DeepFingers, the proposed architecture, produces stable, realistic results that closely follow the physical evolution, whereas ViT introduces nonphysical black islands (hallucinations)  at later times, and DAE-LSTM yields diffused finger tips and spurious  artifacts (hallucinations) at early times. $M$ is the viscosity ratio.
    }
    \label{fig:concentration}
\end{figure*}

The Densenet Autoencoder model (DAE-LSTM)  exhibits noticeable errors in the early stages of flow. Spurious patches of the less viscous fluid (yellow-colored) emerge within otherwise black regions of the more viscous fluid; see Fig.~\ref{fig:concentration} row 4 column 1. At later times, the predicted fingers appear overly smooth, with diffused tips and insufficient splitting. In contrast, the ViT model struggles particularly at later time steps, producing unrealistic structures, such as black islands within yellow regions; see Fig.~\ref{fig:concentration} row 7, column 5. These structures are inconsistent with the underlying physical process. Although these artifacts represent errors in the model’s learning, they are qualitatively distinct from the inaccuracies typically observed in DNS simulations. Such unrealistic flow structures can be regarded as hallucinations of the DL model. These limitations underscore the inherent difficulty of achieving robust generalization across spatial and temporal scales using conventional neural network. In contrast, errors in DNS modeling stem primarily from the accuracy and stability of the spatial and temporal discretizations schemes.

Originally introduced in Neural Machine Translation (NMT)~\cite{lee2018hallucinations}, the term hallucination has since been widely adopted to describe a class of errors commonly observed in Large Language Models (LLMs)~\cite{maynez2020faithfulness}.
However, within the broader AI community, its definition remains vague and inconsistent. As highlighted in a recent review~\cite{maleki2024ai}, no universally accepted characterization or criteria currently exist, and the interpretation of hallucination often varies across domains, sometimes even contradictorily. In general, the term refers to outputs that appear plausible and coherent yet are fundamentally incorrect or unsupported by the input data~\cite{dillion2023can}. AI-generated content, for instance, can produce highly convincing but fabricated or misleading information, raising growing concerns about misinformation and model reliability~\cite{salah2023may}.

In this study, we extend the notion of hallucination to AI models of physics, in particular, physics of porous media flows,
where such behavior manifests as predictions that appear visually coherent but violate fundamental physical laws or deviate from expected flow dynamics. Specifically, we use the term hallucination to describe distinct, physically unrealistic patterns produced by  models, such as the DAE-LSTM and ViT, whose errors cannot be explained by known numerical or DNS-related artifacts. This perspective bridges the conceptual gap between hallucinations in LLMs and spurious behaviors in physics-based DL model, offering a framework to identify, interpret, and mitigate nonphysical predictions in data-driven simulations of fluid dynamics. Overall, the results confirm that while existing architectures such as DAE-LSTM and ViT fail to capture the rich multiscale complexity of VF, DeepFingers succeeds in reproducing the physically realistic evolution of the $c$ field across a range of $M$ and time.

\subsection*{Spectral analysis}

\begin{figure*}
    \centering
    \includegraphics[width=1.0\linewidth]{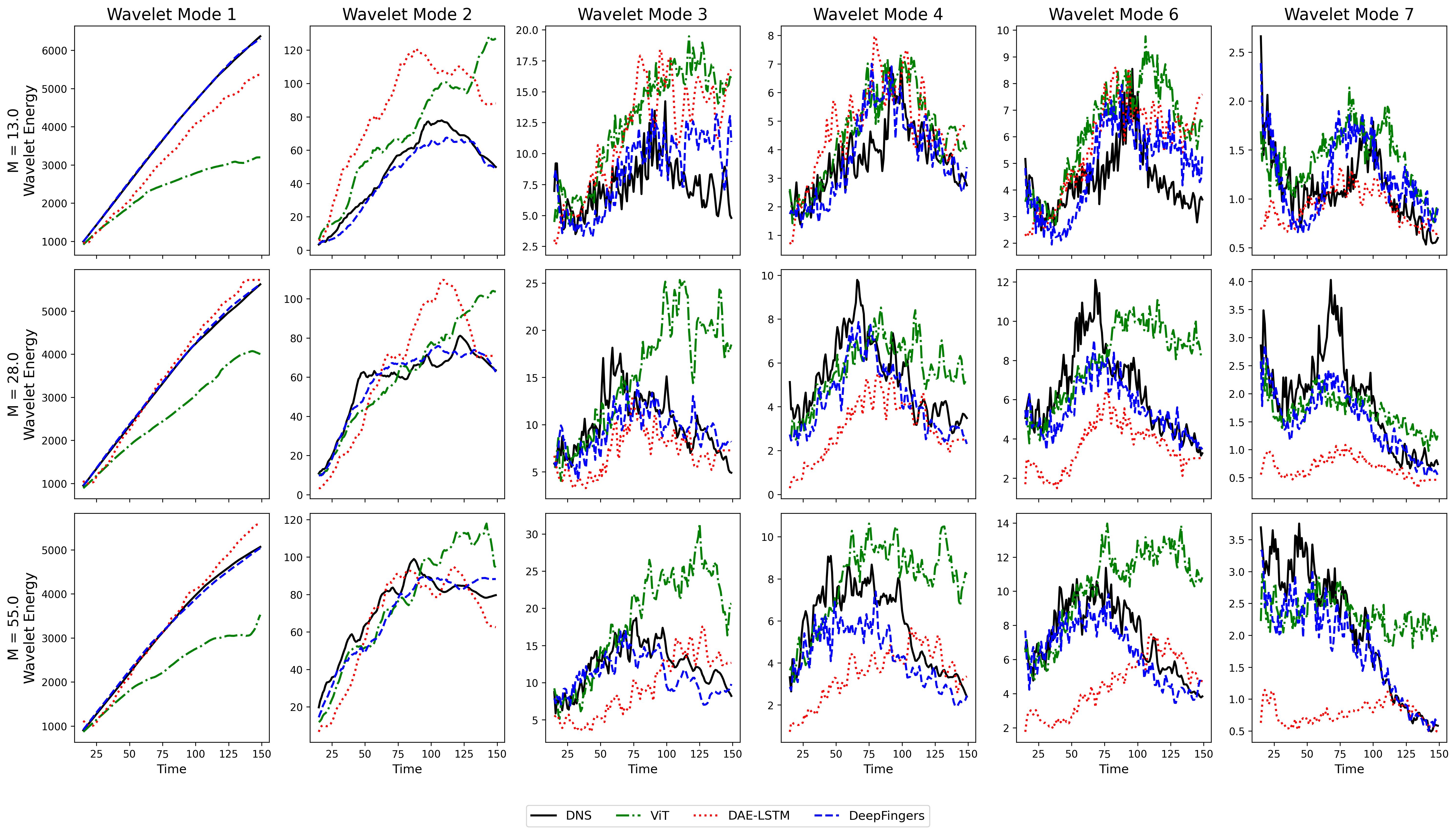}
    \caption{Spectral modes comparison among DNS, DeepFingers, ViT, and DAE–LSTM. 
    DNS shows the expected redistribution of spectral modes from coarse to fine scales. DeepFingers closely follows this evolution, maintaining a balanced distribution across all modes. ViT underestimates the dominant large-scale mode while overestimating higher modes, leading to exaggerated small-scale fluctuations and nonphysical artifacts. DAE–LSTM exhibits $M$ dependent deviations, with higher modes suppressed due to the smaller number of fingers and overly diffused boundaries. Overall, DeepFingers avoids the spectral biases observed in ViT and DAE–LSTM, reproducing the multiscale dynamics more faithfully.
    }
    \label{fig:wavelet plot}
\end{figure*}

Spectral-based mode analysis provides a quantitative framework for assessing the multiscale fidelity of the reconstructed, non-stationary concentration fields of flow. We conduct spectral analysis using wavelets, instead of harmonic functions, because wavelets offer simultaneous localization in time and frequency, making them ideal for analyzing non-stationary, transient signals. The resulting spectrum reveals how features of different scales (incipient fingers to finger merging to channeling) evolve dynamically at each time step. As illustrated in Fig.~\ref{fig:wavelet plot}, the seven wavelet modes exhibit a clear redistribution of spectral energy over time, where lower modes correspond to large-scale, low-frequency structures and higher modes capture small-scale, high-frequency features.

For the ViT model, the first mode remains consistently below DNS, reflecting a slower propagation of the dominant large-scale yellow region shown in Fig.~\ref{fig:concentration}, with this bias later evident in the domain-averaged concentration $\bar{c}(t)$. In contrast, the higher modes (2–7) of the ViT tend to exceed DNS levels, indicating an overestimation of small-scale fluctuations such as fingers, which manifests in mean dissipation rate $\epsilon_c$. 

The DAE–LSTM model exhibits a more complex, $M$-dependent pattern: while mode 1 is lower than DNS only for $M=13$, at higher viscosity ratios ($M=28$ and $M=55$) the higher modes (3–7) fall below DNS. This deviation arises because the model produces fewer fingers overall, leading to weaker fine-scale activity and overly diffused boundaries especially at early time steps. 

In contrast, DeepFingers achieves a balanced distribution across all modes. Although pixel-level finger patterns do not match DNS exactly, an expected consequence of the intrinsic instability of fingering, the modes distribution across scales aligns closely with DNS. This balance indicates that DeepFingers avoids the spectral biases observed in ViT and DAE–LSTM and is therefore better suited in the physics-based evaluations presented in the next section. 

\begin{figure*}
    \centering
    \includegraphics[width=1.0\linewidth]{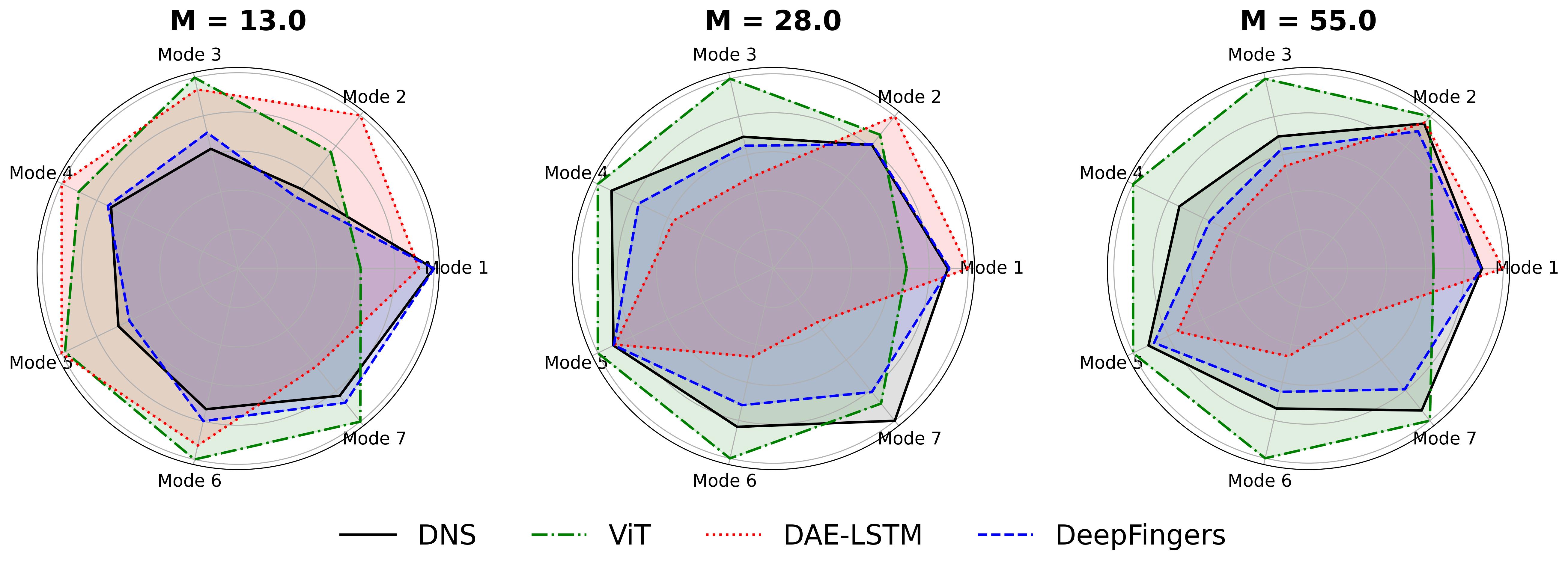}
    \caption{Mean spectral energy comparison among DNS, DeepFingers, ViT, and DAE–LSTM across viscosity ratios $M$, aggregated over all time steps. 
    DeepFingers matches DNS closely across all modes, demonstrating balanced multiscale fidelity. 
    ViT consistently overestimates spectral energy in modes 2–6. In contrast, DAE–LSTM exhibits $M$ dependent behavior: it overestimates energy at $M=13$, but underestimates higher modes at $M=28$ and $M=55$, showing the opposite tendency of ViT due to fewer fingers and overly diffused boundaries.
    }
    \label{fig:wavelet spider plot}
\end{figure*}

To better quantify the overall biases, the mean spectral energy for each viscosity ratio $M$ is calculated by aggregating across all time steps. Fig.~\ref{fig:wavelet spider plot} summarizes these biases for the compared models. DeepFingers aligns closely with DNS across all spectral modes, indicating balanced multiscale fidelity. In contrast, the ViT model consistently produces higher spectral energy in modes 2–6, resulting in a larger overall area relative to DNS. The behavior of the DAE–LSTM model depends on $M$. For $M=13$ the model overestimates energy, but for $M=28$ and $M=55$ the higher modes (3-7) fall below DNS, showing the opposite tendency of ViT by underestimating fine-scale finger structures and producing overly diffused boundaries. 

\subsection*{Addressing hallucination by  spectral debiasing}
The ViT model used in this study relies on the standard attention mechanism of the original architecture, which was primarily designed for static image tasks such as seismic interpretation~\cite{sheng2023seismic}. Consequently, it lacks the capacity to capture temporal dependencies. Extending the architecture to incorporate temporal awareness therefore represents a promising direction for improving predictive performance. In contrast, the CViT architecture~\cite{wang2024cvit} was explicitly developed for spatio-temporal physics problems, employing cross-attention mechanisms tailored to solve dynamic problem. However, its trainable grid-based coordinate embeddings incur substantial memory costs when evaluated over all spatial coordinates during training, necessitating random coordinate sampling. Because fingering dynamics are dominated by fine-scale behavior between two fluids, this sampling strategy prevents CViT from learning the interface evolution effectively, leading to severely blurry result.

\begin{figure*}
    \centering
    \includegraphics[width=1.0\linewidth]{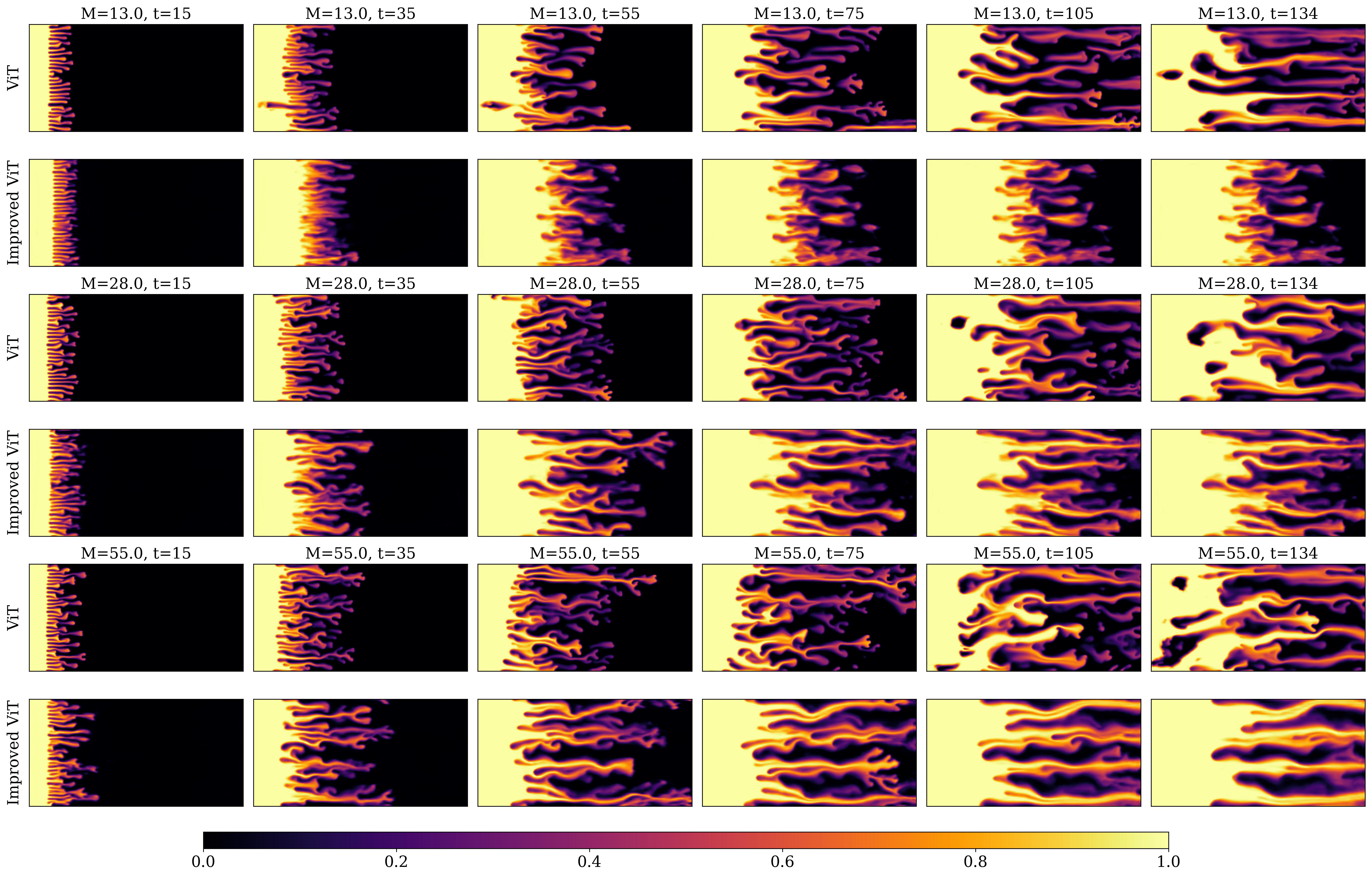}
    \caption{Addressing hallucination in ViT by spectral debiasing.  
    Improved ViT includes cross attention blocks, which removes the nonphysical black (more viscous fluid) islands observed in the original ViT implementation, demonstrating improved temporal prediction and spatial continuity.}
    \label{fig:ViT improved}
\end{figure*}

Motivated by these limitations, we incorporate cross-attention blocks inspired by CViT into the ViT framework. This modification substantially improves temporal coherence and spatial continuity, eliminating artifacts such as the nonphysical black islands previously observed within yellow regions (Fig.~\ref{fig:concentration} row 7, column 5). Nonetheless, challenges remain: at $M = 13$, the reconstructed fingers are still shorter and more diffused than the ground truth, as shown in Fig.~\ref{fig:ViT improved}. These findings underscore both the strong potential of transformer-based architectures for modeling fingering and the need for continued architectural refinement. We further argue that hybrid strategies, such as two-stage frameworks exemplified by DAE-LSTM, represent particularly promising avenues for future research.

\subsection*{Global metrics of fluid mixing from fingering}
We consider four metrics: domain-averaged concentration, breakthrough concentration, degree of mixing, and rate of mixing. See SI for details. 

\subsubsection*{Average concentration}
We compare the temporal evolution of the average concentration $\bar{c}(t)$ from DNS against different AI models---DeepFingers, ViT, and DAE-LSTM---for various viscosity ratio $M$ values. This comparison plot in Fig.~S3 highlights the proposed DeepFingers framework’s ability to capture the dynamics from early to late times. 
The reason is that $\bar{c}(t)$ is strongly influenced by the movement of the less viscous yellow fluid and DeepFingers closely approximates this movement, as shown by 
 Fig.~\ref{fig:concentration}, rows 2, 6, and 10. 
In contrast, the ViT model deviates significantly 
and consistently underestimates the $\bar{c}(t)$ values. Over time, while ViT captures the growth of the fingers, the less viscous fluid regions do not progress as expected. This, combined with the effect of hallucinations manifested as black colored (more viscous fluid) islands of the viscous fluid within the less viscous region causes $\bar{c}(t)$ to remain consistently underestimated. See Fig.~\ref{fig:concentration}, rows 3, 7, and 11, for ViT results. 

For the predictions of DAE-LSTM, the model generally follows the overall trend but exhibits noisy and oscillatory behavior, reflecting limited temporal smoothness and spatial continuity. This indicates that while DAE-LSTM can capture the overall growth of the concentration field, it struggles to maintain smooth spatial evolution over time. 
For $M=13$ (row 4 of Fig.~\ref{fig:concentration}), the slow movement of the less viscous fluid causes $\bar{c}(t)$ to remain consistently below the DNS result. In contrast, for $M=28$ (row 8), the less viscous fluid moves faster than in DNS, leading $\bar{c}(t)$ to exceed the reference values in the later time steps. For $M=55$, DAE-LSTM approximates $\bar{c}(t)$ more accurately, but noise and oscillations persist due to spurious patches of less viscous fluid appearing in the more viscous region at early times and diffused finger tips. Overall, DAE-LSTM struggles to produce consistent performance across different $M$ values.
Table~\ref{tab:global metric results} supports these observations with quantitative evaluation results;  DeepFingers outperforms both ViT and DAE–LSTM. 

\subsubsection*{Breakthrough concentration at outlet boundary}

The concentration breakthrough curve is an important diagnostic of the transport process and is, therefore, often measured in field and lab experiments. 
The DeepFingers model successfully approximates the breakthrough curve across the range of $M$ values (Fig.~S4). Other DL models, ViT and DAE-LSTM, hallucinate in predicting the breakthrough behavior. 
The ViT model predicts too fast breakthrough for the low viscosity contrast, $M=13$, because the ViT fingers grow more rapidly and reach the outlet boundary earlier; see Fig.~\ref{fig:concentration} row 3, column 4. The $M=13$ breakthrough  is faster than the breakthroughs at higher $M$ values (see Fig. S4),  
which is counterintuitive  because the breakthrough time should decrease as the $M$ increases. This is an evidence of hallucination. 
At later time steps for $M=55$, hallucinations in the form of black colored (more viscous fluid) islands within the yellow colored (less viscous fluid) region emerge, while the fingers arriving at the outlet boundary remain thin and exhibit limited growth over time. Consequently, $\bar{c}_{\text{out}}(t)$ tends to flatten, showing a slower increase compared to the other models. This shows that the ViT model lacks  spatial continuity and accurate temporal prediction. 

As for the DAE-LSTM model, a hallucination in the form of a yellow patch within the black region causes a small increase in $\bar{c}_{\text{out}}(t)$ for a short period at an early time step, as shown in 
Fig.~S4 for $M=28$. This suggests that DAE-LSTM also lacks spatial continuity. Similar to the ViT model, for $M=13$ the fingers grow faster and are too long (compared to DNS). For small $M$, the finger growth should be slower due to the dominance of diffusive mechanisms. Consequently, the fingers predicted by DAE-LSTM reach the outlet boundary earlier; see Fig.~\ref{fig:concentration}, row 4, column 4. As $M$ increases, the breakthrough time remains nearly constant, which is counterintuitive. increasing $M$ should result in faster breakthrough, as confirmed by DNS; see Fig.~S4.
This indicates that DAE–LSTM  struggles to produce accurate temporal trajectory predictions. The primary reason is that the latent features generated by the DAE part exhibit a highly complex structure, making it difficult for the LSTM component to learn the corresponding temporal evolution effectively. Simpler latent structures are still captured effectively by DAE during early stages of fingering, when the patterns remain relatively well‑separated and dominated by growth rather than merging or other complex interactions. 



\begin{table}[t!]
\centering
\caption{Global mixing metrics aggregated over all values of the viscosity ratio $M$ in the test dataset. 
RMSE values are relative to DNS. Bold entries indicate the best performance, highlighting DeepFingers as the leading model capable of addressing hallucinations observed in ViT and DAE-LSTM.}
\begin{tabular}{ccccc}
\toprule
Models & \multicolumn{4}{c}{RMSE} \\
\cmidrule(lr){2-5}
 & $\bar{c}$ & $\bar{c}_{\text{out}}$ & $\sigma_c^2$ & $\epsilon_c$ \\
\midrule
\textbf{DeepFingers} & \textbf{0.00410} & \textbf{0.00966} & \textbf{0.00327} & $\mathbf{2.67 \times 10^{-7}}$ \\
ViT                 & 0.11463          & 0.02389          & 0.00523          & $9.10 \times 10^{-7}$ \\
DAE-LSTM            & 0.05334          & 0.03254          & 0.00385          & $8.66 \times 10^{-7}$ \\
\bottomrule
\end{tabular}
\label{tab:global metric results}
\end{table}

\subsection*{Uncertainty propagation and quantification}

\begin{figure}
    \centering
    \includegraphics[width=1.0\linewidth]{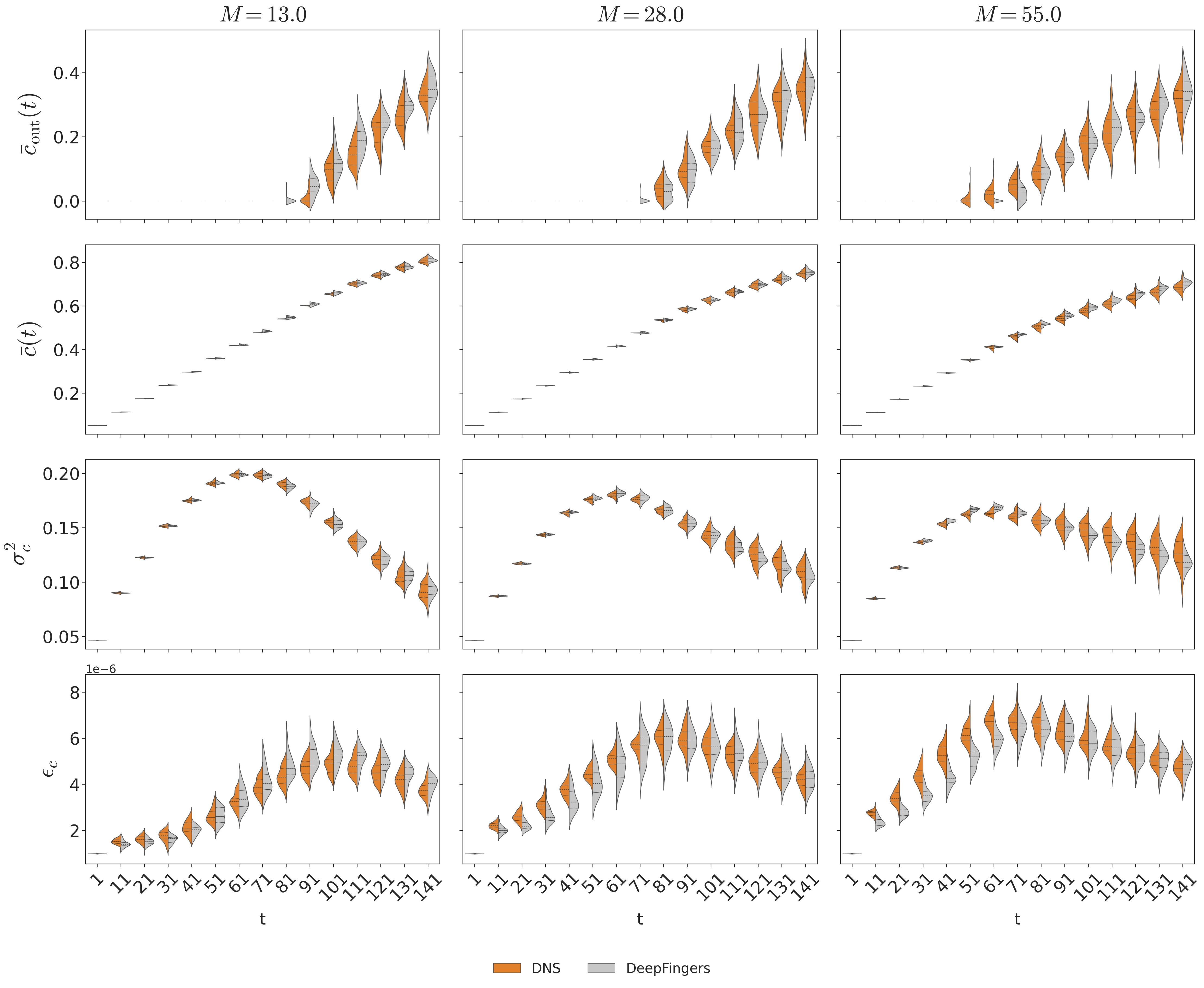}
    \caption{Uncertainty propagation in AI model: time evolution of probability density functions (PDFs) of four mixing metrics $\overline{c}$, $\overline{c}_{\text{out}}$, $\sigma_c^2$, and $\epsilon_c$. Comparison of DeepFingers predictions (gray shading)  with DNS  (orange shading) illustrates that DeepFingers successfully propagates uncertainty. Source of uncertainty is initial condition variability. 
    }
    \label{fig:stochastic}
\end{figure}

In geoscience applications, quantifying uncertainty is essential due to the uncertainty in our knowledge about rock's  heterogeneity and fluid's initial condition. 
In unstable flows, spatial variability in rock properties can significantly influence the initial condition and subsequent flow dynamics. To evaluate how uncertainty impacts fluid mixing and spreading in the domain, we analyze the temporal evolution of probability density functions of the metrics introduced above, as illustrated in Fig.~\ref{fig:stochastic}.
The violin plots demonstrate that DeepFingers provides a reliable approximation of the uncertainty induced by varying initial conditions.

The first three metrics, $\bar{c}(t)$, $\bar{c}_{\text{out}}(t)$, and $\sigma_c^2$ show that DeepFingers closely captures both the variance and median behavior observed. 
For the $\epsilon_c$, at $M=28$ during early time steps, DeepFingers exhibits a slightly higher variance than DNS, but converges toward the DNS distribution as time progresses. At $M=55$, DeepFingers initially underestimates the distribution relative to DNS, yet aligns more closely at later time steps.

As $M$ increases, the system exhibits more complex fingering behavior. Early time steps are dominated by advection, leading to dynamic and unstable finger growth, while later stages are increasingly governed by diffusion, resulting in smoother and more predictable patterns. Such fingering-driven uncertainty is visible even when the initial condition is certain, as shown by Fig.~S2. 
The phenomenon exhibits dynamics that bear resemblance to chaotic and turbulent behavior. 

\section*{Discussion}\label{sec12}

AI has been increasingly positioned as a transformative tool for physics, with ambitious claims ranging from accelerated discovery to the replacement of direct numerical simulation in complex flow systems. Deep Learning (DL) models offer a powerful alternative to traditional numerical solvers for approximating complex flow physics directly from data. This approach addresses long-standing challenges in simulating multiscale, nonlinear processes such as viscous fingering, where conventional methods can be computationally expensive or become numerically unstable at high values of the viscosity ratio or Peclet number.
We show that state-of-the-art AI models can generate predictions that are visually plausible yet fundamentally nonphysical, violating governing principles of fluid motion. We formalize this phenomenon by extending the concept of hallucination from large language models (LLMs) to physics-based AI, thereby providing a unifying framework to interpret spurious yet convincing predictions in data-driven simulations. 

While hallucination has been widely discussed in the context of natural language and computer vision models, its appearance in scientific modeling remains unexplored. Our finding has significant implications for both the foundations of AI and the reliability of data-driven modeling in the physical sciences.  This conceptual bridge clarifies that hallucination is not domain-specific, but instead reflects deeper inductive biases of learning architectures when applied to multiscale physical systems.
By integrating the Fourier neural operator within a deep operator network to address spectral bias, we design a new AI architecture for fingering dynamics that can not only replicate the intricate spatio-temporal evolution of fingering but also adhere to key underlying physics that govern fluid mixing. 

Another major contribution is the explicit treatment of uncertainty, which lies at the core of all subsurface processes because  natural heterogeneity lead to inherently stochastic flow behaviors. Conventional AI-based surrogate models often overlook this aspect, producing single-point estimates that fail to represent the range of plausible outcomes. 
We show how to leverage the intrinsic stochasticity to generate a sets of physically consistent solutions, thereby enabling the propagation of uncertainty in both space and time. This capability is especially significant for decision-making in reservoir management and other geoscientific applications, where understanding the variability of outcomes is crucial.  


\section*{Methods}\label{sec11}
\subsubsection*{Fourier Neural Operator (FNO)}

The goal of the FNO is to learn the operator $\mathcal{G}: A \mapsto B$. Here, $A$ denotes the input function and $B$ the corresponding output function:
\begin{equation} \label{eq:ab_mapping}
A : r \mapsto A(r), \quad 
B : r \mapsto B(r), \quad r \in D
\end{equation}
where $r$ represents a point in the domain $D \subset \mathbb{R}^d$ and may correspond to a spatial coordinate and can also include time if the problem is spatio-temporal, e.g., in viscous fingering. Thus, both $A(r)$ and $B(r)$ are functions defined over the same domain $D$. 

The original implementation~\cite{li2020Fourier} introduces the operator layer as an iterative framework 
to learn the mapping from an input function $A$ to an output function $B$:

\begin{equation} \label{eq:operator_composition}
B = \mathcal{G}(A) 
   = \bigl( \mathcal{Q} \circ \mathcal{L}^{(L)} 
   \circ \cdots \circ \mathcal{L}^{(1)} \circ \mathcal{P} \bigr)(A)
\end{equation}
Here, $\circ$ denotes function composition, and $L$ represents the number of layers. 
The operator $\mathcal{P}$ lifts the input into the initial latent representation $z^{(0)}$ through a linear layer, 
while each $\mathcal{L}^{(\ell)}$ corresponds to the $\ell$-th non-linear operator layer. 
The operator $\mathcal{Q}$ projects the final latent representation $z^{(L)}$ to obtain the output function $B$. 
The non-linear operator defined in Equation~\eqref{eq:operator_composition} 
updates the representation $z^{(\ell)} \mapsto z^{(\ell+1)}$ as follows:

\begin{equation} \label{eq:layer_definition}
\mathcal{L}^{(\ell)}\!\left(z^{(\ell)}\right) 
= \sigma \!\left(W^{(\ell)} z^{(\ell)} + K^{(\ell)}\!\left(z^{(\ell)}\right)\right)
\end{equation}
where $\sigma$ is a non-linear activation function applied component-wise, 
$W^{(\ell)}$ denotes a linear transformation, 
and $K^{(\ell)}$ represents a kernel integral operator defined via the Fourier transform. 
The kernel operator is computed as:

\begin{equation}\label{eq:kernel_integral_operator}
K^{(\ell)}\!\left(z^{(\ell)}\right) 
= \mathcal{F}^{-1}\!\left( R^{(\ell)} \cdot \mathcal{F}\!\left(z^{(\ell)}\right)\right)
\end{equation}
where $\mathcal{F}$ and $\mathcal{F}^{-1}$ denote the Fourier and inverse Fourier transforms, respectively, 
and $R^{(\ell)}$ is a complex-valued weight tensor applied to a truncated set of Fourier modes.




The FNO was further extended to U-FNO~\cite{wen2022u}, which improves the preservation of multi-scale features by concatenating fine and coarse level representations across layers. This is essential for modeling fingering, which is known to display spatial features over a range of length scales, from tiny incipient fingers at the pore-scale to  long channels at the domain-scale. 
In particular, U-FNO is effective at retaining high-frequency Fourier modes, which are essential for capturing fine-scale details 
such as finger structures. Fig.~\ref{fig:FNO-deepOnet} illustrates 
the key differences between FNO and U-FNO. 

Compared to the standard FNO, U-FNO incorporates an additional U-Net based convolutional operator, denoted by $U^{(\ell)}$. Let $h^{(\ell)}$ is the representation feature that propagates through and is updated by the U-FNO layers. We define a new U-FNO operator layer $\mathcal{U}^{(\ell)}$, which updates the representation 
$h^{(\ell)} \mapsto h^{(\ell+1)}$ as follows:

\begin{equation} \label{eq:unet_layer_definition}
\mathcal{U}^{(\ell)}\!\left(h^{(\ell)}\right) 
= \sigma \!\left(W^{(\ell)} h^{(\ell)} 
+ U^{(\ell)}\!\left(h^{(\ell)}\right) 
+ K^{(\ell)}\!\left(h^{(\ell)}\right)\right)
\end{equation}
where $U^{(\ell)}$ introduces multi-scale convolutional features, while $K^{(\ell)}$ and $W^{(\ell)}$ follow the definitions in Equation~\eqref{eq:layer_definition}. By combining these components, the network 
constructs a richer representation that is better suited to capturing underlying physics of multi-scale dynamics and complex interactions of VF.






\subsubsection*{Deep Operator Networks (DeepONet)}

DeepONet approximates a nonlinear operator $\mathcal{G}: A \mapsto B$ by mapping between two 
infinite-dimensional function spaces using two separate networks, namely the branch and trunk networks. 
The branch network processes the input function $A$, while the trunk network encodes the query location 
$y \in Y$. Together, they approximate the output function $\mathcal{G}(A)(y)$ or $B$.

To be precise, let $(r_1, \dots, r_m)$ denote a collection of $m$ grid points used to discretize the 
input function $A$ expressed as $A_m = \bigl(A(r_1), \dots, A(r_m)\bigr)$.
The branch network transforms the discretized input $A_m$ into a vector of branch outputs
$b = (b_1, \dots, b_q) \in \mathbb{R}^q$. Simultaneously, the trunk network evaluates the query location 
$y \in Y$ to produce the associated trunk outputs
$\varphi = \bigl(\varphi_1, \dots, \varphi_q\bigr) \in \mathbb{R}^q$.
The output of DeepONet is obtained by combining the branch outputs with the trunk outputs through a merging operator.
\begin{equation}\label{eq:don_dot}
B = \mathcal{G}(A_m)(y) = \sum_{i=1}^q b_i(A_m) \, \varphi_i(y)
\end{equation}
This formulation combines the two representations and allows DeepONet to generalize across different function spaces and spatial domains, with theoretical guarantees of universal approximation~\cite{lu2019deeponet}. 

\subsubsection*{DeepFingers}

The proposed DeepFingers architecture, as shown in Fig.~\ref{fig:FNO-deepOnet}, introduces a hybrid framework for modeling the fingering problem by combining the strengths of DeepONet and FNO. While retaining the general structure of DeepONet, the branch network is augmented with a single FNO layer ($L=1$), as defined in Equation~\eqref{eq:operator_composition}, to efficiently capture global, non-local dependencies of the input function. However, we remove the operator $\mathcal{Q}$ in Equation~\eqref{eq:operator_composition}. The outputs of the branch and trunk networks are subsequently merged and refined through two consecutive U-FNO layers, which are essential for extracting multi-scale features and resolving the sharp interfacial patterns characteristic of fingering dynamics. A final projection layer $\mathcal{S}$ is applied to produce a single-channel output solution in Fig.~\ref{fig:FNO-deepOnet}.

DeepFingers takes the concentration field $c_t$ as the input function, denoted by $A$ in Equations~\eqref{eq:operator_composition} and \eqref{eq:don_dot}. 
In the trunk network, we define the parameter vector $\xi$, consisting of two scalar values: 
the viscosity ratio $M$ and the future time step $(t+1)$. Formally,
\[
\xi = [M, t+1]
\]
where $\xi$ corresponds to $y$ in Equation~\eqref{eq:don_dot}. Accordingly, the VF problem addressed by DeepFingers is formulated as learning the operator mapping:
\[
\mathcal{G} : (c_t, \xi) \mapsto c_{t+1}
\]
Both the input function $c_t$ and the output function $c_{t+1}$ are discretized on a cartesian two-dimensional grid of shape \( (n_y', n_x') = (64, 128) \) representing the spatial domain of the concentration field, channel $C_{in}=1$. Thus, the $c_t$ and $c_{t+1}$ tensors have a shape of $(1, 64, 128)$.

Let $b_{out}$ and $t_{out}$ are the output of branch and trunk networks respectively. Both are lifted to produce output channel of $C_{out} = 64$ thru operator $\mathcal{P}$ in FNO layer of branch network and linear transformation in trunk network. The results are merged thru pointwise multiplication operation as indicated by crossed node in Fig.~\ref{fig:FNO-deepOnet}.
\begin{equation}\label{eq: merger_operation}
h^{(0)} = b_{out} \odot t_{out}
\end{equation}
where $h^{(0)}$ has shape $(C_{\text{out}}, n_y', n_x') = (64,\, 64,\, 128)$. Representation $h^{(0)}$ is the initial representation $h$ before updated in Equation~\eqref{eq:unet_layer_definition}. We employ a series of two U-FNO layers to update representation $h^{(0)}$ sequentially according to Equation~\eqref{eq:unet_layer_definition}:

\[
\begin{aligned}
h^{(1)} &= \sigma \!\left( 
        K^{(1)}(h^{(0)})
        + U^{(1)}(h^{(0)}) + W^{(1)} h^{(0)}
\right) \\[6pt]
h^{(2)} &= \sigma \!\left( 
        K^{(2)}(h^{(1)})
        + U^{(2)}(h^{(1)}) + W^{(2)} h^{(1)}
\right) \\[6pt]
\end{aligned}
\]
Lastly, the operator $\mathcal{S}$ projects $h^{(2)}$ back to a single‑channel field, ensuring that the output dimensionality matches $C_{in}=1$.
\[
\begin{aligned}
c_{t+1} &= \mathcal{S}\!\left( h^{(2)}
\right) \\[6pt]
\end{aligned}
\]
This projection completes the operator mapping by producing the predicted concentration field at the next time step $c_{t+1}$.

\section*{Funding Declaration}
R.W. acknowledges funding support from the USC Don Paul Fellowship. B.J. acknowledges funding support from the Aramco Research Project.

\section*{Author Contribution}
B.J. designed the study and created the DNS model.  R.W. created and trained the AI models and prepared the plots. R.W. and B.J. interpreted and analyzed the plots. R.W. and B.J. wrote the initial manuscript draft. B.J.  wrote the final manuscript.

\backmatter

\bibliography{FullDomain}

\end{document}